\documentclass[prd,11pt]{revtex4-1}

\usepackage{bm}
\usepackage{amsmath}    
\usepackage{graphicx}   
\usepackage{verbatim}   
\usepackage{color}      
\usepackage{subfigure}  
\usepackage{hyperref}   
\definecolor{amaranth}{rgb}{0.9, 0.17, 0.31}
\definecolor{purple(munsell)}{rgb}{0.62, 0.0, 0.77}
\definecolor{americanrose}{rgb}{1.0, 0.01, 0.24}
\definecolor{palatinateblue}{rgb}{0.15, 0.23, 0.89}
\definecolor{royalblue(web)}{rgb}{0.25, 0.41, 0.88}
\definecolor{hanpurple}{rgb}{0.32, 0.09, 0.98}
\definecolor{beaublue}{rgb}{0.74, 0.83, 0.9}
\definecolor{carminered}{rgb}{1.0, 0.0, 0.22}
\definecolor{brightpink}{rgb}{1.0, 0.0, 0.5}
\definecolor{vividviolet}{rgb}{0.62, 0.0, 1.0}

\hypersetup{ linktoc=all,
    colorlinks, linkcolor={palatinateblue},
    citecolor={brightpink}, urlcolor={amaranth}}

\newcommand{\be}{\begin{equation}}
\newcommand{\ee}{\end{equation}}
\newcommand{\bs}{\begin{split}} 
\newcommand{\bea}{\begin{eqnarray}}
\newcommand{\eea}{\end{eqnarray}}

\begin{document}

\title{Evolution of the concept of the curvature in the momentum space   } 

\author{Nosratollah Jafari}\email{nosrat.jafari@fai.kz}

\affiliation{Fesenkov Astrophysical Institute, 050020, Almaty, Kazakhstan;
\\
Al-Farabi Kazakh National University, Al-Farabi av. 71, 050040 Almaty, Kazakhstan;
\\
 Center for Theoretical Physics, Khazar University, 41 Mehseti Street, Baku, AZ1096, Azerbaijan}


\begin{abstract}

We review the history of the curved momentum space from Max Born to the recent relative locality, $\kappa$-Poincare and noncommutative geometric proposals. We found that the concept of the curvature in the momentum space and motivations has been evolved during 80 years from the Max Born time. Motivations has been evolved from introducing general relativistic like equation for the momentum space to the relaxation of the concept of locality in special relativity and non-commutativity. This study can help us for a better understanding of this concept in quantum gravity. 

\end{abstract}

\maketitle

\tableofcontents

\section{Introduction } \label{sec1}

This is an investigation about the history and concept of the  introducing curvature in the momentum space. The concept of this curvature has evolved during 80 years after introducing firstly by Max Born \cite{Born1938ASF}. There are different proposals for introducing curvature to the momentum space. Born treated the momentum space as a normal tangent space to the position space. Russian physicists  introduced curvature into momentum space for removing infinities in the field theory before renormalization methods\cite{Golfand:1962kjf}. E. R. Caianiello  tried to express quantum mechanics as a pure geometric theory like general relativity in the gravity case\cite{caianiello1980geometry}. Shahn Majid has related curvature in the momentum space to the noncomutattative geometry \cite{Majid:1988we}. Relative locality proposal has been removed locality in the special relativity and given some definitions for the metric, curvature and torsion of the momentum space \cite{Amelino-Camelia:2011lvm}. We can generalize the Hamiltonian equations in the 8-dimensional curved phase space\cite{Barcaroli:2015xda}. Finally, $\kappa$-Poincare  has been involved curved momentum space as a natural generalization of special relativity \cite{Kowalski-Glikman:2013rxa}. In the following sections we investigate these proposals in details. Throughout we use natural units, $\hbar=c=1$.

\section{Born proposal } \label{sec2}

The first suggestion for a curved momentum space was given at the third decade of the past century by Max Born \cite{Born1938ASF}. The \emph{principal of reciprocity} states that the laws of nature are invariant under  
\be    x_\mu   \rightarrow p_\mu ~~,~~~~  p_\mu   \rightarrow -x_\mu.\ee
Born has obtained this reciprocity principle by considering the motion of a free particle in quantum mechanics which is represented by a plane wave
 \be \psi(x)= C \exp(ix_\mu p^\mu/\hbar),\ee
here C is a constant. Any wavefunction  in the spacetime can be transformed to a wavefunction  in the momentum space by a Fourier transform
 \be \phi(p)= \int \psi(x)\exp(ix_\mu p^\mu/\hbar) dx. \ee
These wavefunctions are symmetric in x and p, and this symmetry is the essence of the Born reciprocity.

We denote this duality map with I and it constitutes a complex structure in the phase space such that $I^2= -1$. The kinematical symmetry generated by $I$ is broken with gravity. In general relativity space-time is curved, while energy-momentum space is flat.
Born argued that the momentum space should also be a curved space similar to the spacetime in the Einstein's general relativity. In fact, quantum mechanics forces us to consider phase space as the natural arena for physics.

For the metric in the momentum space he discussed that if the laws of the nature are classical mechanics then transformations laws for the momentum $p_\mu$ will be determined solely from the transformations for $x^\mu$. Then, assuming an independent quadratic form for line element in the momentum space would not be reasonable. But, the laws of the nature should be the quantum mechanics laws and we have some rooms for independent line element in the momentum space. For the larger bodies and a limited energy we have line element in the spacetime  and for the small region in the spacetime and an unlimited large amount of the energy we have a line element in the momentum space. For the line element in the momentum space we write 
\be ds^2= h^{\mu \nu} dp_\mu dp_\nu,             \ee
where $ h^{\mu \nu}$ is the metric of the momentum space.
Born proposed an equation similar to Einstein's general relativity 
\be       R_{\mu \nu}  - \frac{1}{2} R g_{\mu \nu} = - 8 \pi G T_{\mu \nu}, \ee
in the momentum space
\be       \Tilde{P}^{\mu \nu}  - \frac{1}{2}  \Tilde{P} h^{\mu \nu} = - \Tilde{k} \Tilde{T}^{\mu \nu},\ee
where $\Tilde{P}^{\mu \nu}$ is the Riemann curvature tensor in the momentum space and $\Tilde{P}$ is the corresponding Ricci scalar curvature. Also, 
$ \Tilde{T}^{\mu \nu}$  is the spacetime correspondence of the energy-momentum tensor $T_{\mu \nu}$, and $\Tilde{k} $ is a constant. 
We can interpret
\be    \int  \Tilde{T}^{\mu 0} dp_x dp_y dp_z \equiv x^\mu, \ee
as space coordinates and time value of the system. But, Born's suggestion has not lead to good results for the quantization of gravity as discussed by Amelino-Camelia \cite{Amelino-Camelia:2012vpb}. 

The next attention to the curved momentum space was due to the 1960-1990 period and especially from Russian physicists  \cite{Golfand:1962kjf}. They were looking for a divergence free quantum field theory. 

\section{ Caianiello's proposal} \label{sec3}

The next attention to the curved momentum space was due to the 1960-1990 period and especially from Russian physicists  \cite{Golfand:1962kjf}. They were looking for a divergence free quantum field theory.

Around 1980 Caianiello and his collaborators interpreted quantization as curvature in the  8-dimensional spacetime tangent bundle \cite{caianiello1980geometry, Caianiello:1985nv}. In fact, they tried to express quantum mechanics as a pure geometric theory like general relativity in the gravity case . In this approach momentum space is a curved space  which has been immersed in these 8-dimensional phase space. They considered an Hermitian metric tensor $G=\{g_{\mu\nu} \}$ as
\be  G =  G'+ iG'', \ee
the geometric vectors at the point $ \textbf{P}\equiv \textbf{x}$ are denoted by
 \be V=\{V^\mu \},~~~~~ V^\mu= V'^\mu+ i V''^\mu . \ee
A displacement by a change of gauge on $V^\mu$ will get 
 \be d V^\nu= \Gamma_\mu dx^\mu V^\nu,  \ee
 which moves $\textbf{P}$ on the manifold. This require a change   
\be  V^\nu\rightarrow V^\nu(1+ \Phi_\mu dx^\mu ),          \ee
in which $\Phi_\mu$ is a gauge $\Phi_\mu$ connection. Thus, the connection in general can be written as 
\be    \Gamma_\mu= \Tilde{\Gamma}_\mu + \Phi_\mu , \ee
for which we have two limiting case:\\
1. All $\Gamma =\Tilde{\Gamma}$ are real, and $ \Phi_\mu=0 $,

in this case we obtain usual standard general relativity with the standard connection $\Tilde{\Gamma}\mu $.\\ 
2. All  $\Gamma$ are skew Hermitian,  
    $\Gamma_\mu=-\frac{i}{\hbar} F_\mu$ , and $ \Phi_\mu= \frac{i}{\hbar} A_\mu $,
    
in which $ F_\mu= F^\dag_\mu   $ and $ A_\mu= A^\dag_\mu   $, and the Caianiello's model is built for this case.\\ 
Quantum mechanics  conditions for a free particle are 
\be [Q_\mu, Q_\nu] = 0,~~~~  [P_\mu, P_\nu] = 0,~~~~  [Q_\mu, P_\nu]=i\hbar \delta_{rs}. \ee 
By setting
\be P_\mu= -i \hbar D_\mu= -i \hbar \frac{\partial}{\partial x^\mu} + F_\mu, ~~~~ 
Q_{\Tilde{\mu}} = -i \hbar D_{\Tilde{\mu}} = -i \hbar \frac{\partial}{\partial p_{\Tilde{\mu}}} - F_{\Tilde{\mu}}, \ee
in which $\Tilde{\mu}= \mu +4 $ shows indices for the momentum space, we find 
\be [D_{\Tilde{\mu}}, D_{\Tilde{\nu}}] = 0,~~~~  [D_\mu, D_\nu] = 0,~~~~  
[D_{\Tilde{\mu}}, D_\nu]=i\hbar \delta_{{\Tilde{\mu}} \nu}. \ee

We can take vector $ V^I = \psi^I $ as a vector in the Hilbert space, where $ I,J,...=0,...,7 $ are indices of the eight dimensional phase space. Since the tensorial or spinorial character is not changed by the application of a quantum mechanics  operator on it, the standard covariant derivative will convert to a absolute derivative along the $\mu$-axis in the quantum frame. If we write   

\be      \psi^I_{;J}= D_J\psi^I,      \ee 
then we can compute      
\be  [D_I, D_J] \psi^K = R^K_{~~L I J}\psi^L,  \ee 
in which $R^\lambda_{~~\rho \mu \nu} $ is the Riemann curvature tensor. Thus, quantization appears here as the curvature in the phase space. 
Here, quantum mechanics imposes constraints  on the commutators of the coordinates and momentums, and these commutators determine the curvature tensor of the phase space.

In  Caianielloi's approache spacetime and the momentum space line elements are combined into one line element
\be  \label{8dimMetric} dS^2 = dt^2 - dx^2 - dy^2 - dz^2 + \frac{1}{b^2}( dE^2 - dp_x^2 - dp_y^2 - dp_z^2 ), \ee
in which $b$ is a constant that depends on the Planck scale  \cite{Caianiello:1982zz, Caianiello:1989wm}.
In terms of the modified metric we can write
\be \Tilde{g}_{\mu \nu}= \Big( 1 - c^4\frac{\ddot{x}^\rho \ddot{x}_\rho}{a^2_{max}} \Big) g_{\mu \nu}, \ee
in which $a_{max}$ is the maximal acceleration and it has been studied extensively \cite{Caianiello:1981jq}.
The causality constraint implies that proper accelerations must be limited
 \be |\ddot{x}| \leq  a_{max}, \ee
and from the Heisenberg uncertainty relation we found that
 \be a_{max}= \frac{2mc^3}{\hbar}.  \ee

We can also look for the most general transformations in the 8\textbf{D} phase space which leaves this line element as an invariant. The group of transformations is $ U(1,3 )$. For this line element the Minkowski and Newtonian line elements are limiting forms, just as the Newtonian spacetime is a limit of the Minkowski spacetime. 
Maximal acceleration will give us a generalized uncertainty principle   
\be    [x^\mu, p^\nu]=i \hbar \Big( 1 - c^4\frac{{\ddot{x}^\rho \ddot{x}_\rho}}{a^2_{max}} \Big)^{-1} \eta_{\mu \nu},\ee
 which is an essences for the quantum gravity phenomenology \cite{Capozziello:1999wx}. Recently, Castro and others have used this 8-dimensional line element and the Born reciprocity for extending general relativity \cite{Slawianowski, Low1993U31TW, Castro:2008zzc, Pavsic:2009xv, CastroPerelman:2020ysg}. 

Quantum geometry has emerged from this 8-dimensional line element  Eq.~(\ref{8dimMetric}) as space-time tangent bundle $ M_8 = V_4 \otimes T V_4 $, in
which $V_4$ is space–time with metric $g_{ij}$ and $TV_4$ is momentum space with metric $ h^{\mu \nu}$, and here we put $i,j,..=0,...,3$ and $\,u \nu, ...=4,...,7$. If we write $dX_A= [dx^i, dp_\mu]$ in which $A, B=0,...,7$, then by using the phase space metric
\be  G_{AB}=\begin{bmatrix}
 g_{ij} & 0 \\
0 & h^{\mu \nu} 
\end{bmatrix}  \ee

we can write line element in this 8-dimensional phase space as
\be   dS^2= G_{AB} dX_A dX_B= g_{ij}dx^i dx^j + h^{\mu \nu} dp_\mu dp_\nu. \ee
Then by introducing the Einstein's tensor $\Tilde{G}_{AB}$ and the Einestein field equations will take  a block-diagonal form as
\be  \Tilde{G}_{AB} =\begin{bmatrix}
 \Tilde{G}_{ij} & 0 \\
0 & \Tilde{H}^{\mu \nu} 
\end{bmatrix}  =
\begin{bmatrix}  
k T_{ij} & 0 \\
0 & \Tilde{k} \Tilde{T}^{\mu \nu} 
\end{bmatrix}
\ee

for the Born's conjecture. 

\section{ Majid's proposal and Relative Locality} \label{sec5}

Around the beginning of the this century new attentions have been grown to the curved momentum space. Some of the mathematical and 
the conceptual papers by Shahn Majid dated from the late 1980 decade have discussed the necessity for introducing curvature in the momentum space \cite{Majid:1988we, Majid:1999tc, Majid:1999td}. He has argued that the curvature in the position space implies non-commutativity in the momentum space. As an example when the position space is a 3-sphere which have momentum algebra $su2$, then the  enveloping algebra $U(su2)$ will be satisfied in  
\be [p_i, p_j ] = \frac{c_1}{R}\epsilon_{ijk} p_k, \ee
in which $R$ is related to the radius of the curvature of the 3-sphere and $c_1$ is a constant. By using the Born reciprocity then we have possibility for the curvature in the momentum space. 
In Majid's approach the curved momentum space is alongside with the noncomutativity in the spacetime \cite{Majid:1999tc}. 

In 2011 the relative locality theories, which are based on the nonlinear combinations of momenta, were introduced by G. Amelino-Camelia and his colleagues\cite{Amelino-Camelia:2011lvm,Amelino-Camelia:2011hjg}. In these theories, momentum space $ \textsf{P}$ is assumed to have a nontrivial geometry, and we give some definitions for the connection, curvature and torsion in the momentum space manifold. The addition rule is defined by a $C^{\infty} $ map
\be  \label{Addition}  \oplus: \textsf{P} \times \textsf{P} \longrightarrow \textsf{P}, ~~~~
 (p,q)\rightarrow p \oplus q, \ee
 
For this combination rule one could define a connection by
\be \Gamma_{c}^{ab}(0)=-\frac{\partial}{\partial p_a} \frac{\partial}{\partial q_b}
(p \oplus q)_c|_{q,p=0}.\label{connection}\ee
The torsion of this combination is defined by,
\be T_c^{ab}(0)=-\frac{\partial}{\partial p_a}
\frac{\partial}{\partial q_b}[(p\oplus q)_c-(q\oplus p)_c]|_{q,p=0}.\label{torsion}\ee
Using this connection, Eq.~(\ref{connection}), one can define the \emph{curvature} of the 4-momentum space,
\be \label{curvture} R^{abc}_{~~~d}(0) = 2\frac{\partial}{\partial p_{[a}} \frac{\partial}{\partial q_{b]}}
\frac{\partial}{\partial k_c}\Big((p\oplus q)\oplus k-p\oplus(q\oplus
k)\Big)_d\Big|_{p,q,k=0}, \ee
where as usual, the bracket denotes anti-symmetrization. Curvature could be interpreted as a lack of associativity of the combination rule. The mass of a particle is interpreted as the geodesic distance from
the origin of the momentum space. Different observers see different spacetimes which means coincidences of events are not the same for all observers.

\section{  Hamiltonian mechanics with a curved momentum space} \label{sec6}

 In a similar way, we can generalize the Hamiltonian equations in the 8 dimensional curved phase space  \cite{Barcaroli:2015xda, Relancio:2020rys}. This approach belong to the Hamiltonian geometry category which is also related to the Finsler geometry   \cite{Pfeifer:2019wus}. Finsler geometry is a generalization of Riemann geometry with a position and velocity dependent metric.

Starting from the free particle dispersion relation 
\be E^2= \textbf{p}^2 + m^2, \ee
and defining Hamiltonian as 
\be    H(x, p) = m^2,      \ee
we can introduce deformations as 
 \be H(x, p) = p_0^2 - \textbf{p}^2 + l_p Q^{abc} p_a p_b p_c, \ee
in which $l_p$ is the Planck length and $ Q^{abc}$ is a matrix of numerical coefficients.

In Mathematics, the Hamiltonian is a function on the cotangent bundle. This cotangent bundle is 
cotangent to the a spacetime manifold. We define phase space metric as
\be  g^H_{\mu \nu} (x,p)= \frac{1}{2} \frac{\partial}{\partial p_\mu} \frac{\partial}{\partial p_\nu} H(x, p),  \ee
from which and using a non-linear connection we can find the curvature of the momentum space \cite{Barcaroli:2015xda}.

\section{ kappa-Poincar\`e  and curved momentum space} \label{sec7}

In the first years of the twenty century it was shown that the DSR and the $\kappa$-Poincare theories can be understood in terms of the curved momentum space as a subspace of the de Sitter space \cite{Kowalski-Glikman:2003qjp}. For continuation we consider the momentum space of a $\kappa$-Poincar\`e particle which is a four dimensional group manifold of a Lie group $AN(3)$ 
\cite{Kowalski-Glikman:2013rxa}. The generators  of the Lie algebra for this group satisfy 
 \be [X^0,X^i]= \frac{i}{\kappa} X^i, \ee 
  Matrix representations of this Lie algebra are 5-dimensional ones which one of them is the abelian generator $X^0$ and the others are three nilpotent generators $X^i$. After exponentiation we get the 5-dimensional Minkowski space with the coordinates $(P_0, P_i, P_4)$ 
\be    P_0 = \kappa \sinh{\frac{p_0}{\kappa}}  + \frac{\textbf{p}^2}{2\kappa} e^{p_0/\kappa}, ~~~ P_i = p_i e^{p_0/\kappa},
~~~    P_4 = \kappa \cosh{\frac{p_0}{\kappa}}  - \frac{\textbf{p}^2}{2\kappa} e^{p_0/\kappa},\ee
which satisfy
\be - P_0^2 + \textbf{P}^2 + P_4^2 = \kappa^2, \ee 
the de Sitter space condition and $p_\mu$ are the flat coordinates. The line element in the flat coordinates will be
\be  \label{KPionMetric} ds^2 = -dp_0^2 + e^{2p_0/\kappa} d\textbf{p}^2.\ee
For this metric we can calculate the distance function
\be C(p) = \kappa^2 \cosh^{-1}{\frac{P_4}{\kappa}}.\ee  
Thus, the mass-shell condition will be 
\be \cosh{\frac{p_0}{\kappa} - \frac{\textbf{p}^2}{2\kappa} e^{p_0/\kappa}}=\cosh{\frac{m}{\kappa}}.
\ee

In a close similarity, we have $\kappa$-Minkowski deformations which are obtained from the deformed coordinates 
 \be     [x_i, x_j]=0, ~~~ [x_0, x_i]= \frac{i}{\kappa} x_i. \ee
The geometrical structure of the $\kappa$-Minkowski momentum space is interesting \cite{Lizzi:2020tci, Juric:2012xt, Juric:2013mma}. For example, combinations of the plane waves in these non-commutative coordinate will be given in a nontrivial way
 \be \exp{(ik_\mu x_\mu)}\exp{(iq_\mu x_\mu)}=  \exp{ \Big\{ i\frac{(k_0 + q_0)/\kappa}{e^{(k_0+ q_0)/\kappa} - 1}  
  \Big[ \Big(\frac{e^{k_0/\kappa} - 1}{k_0/\kappa} \Big) k_i + e^{-k_0/\kappa} \Big(\frac{e^{k_0/\kappa} - 1}
  {q_0/\kappa} \Big)q_i  \Big]x^i  + i(k_0 + q_0)x^0 \Big\} }.\ee
 This type of non-triviality shows that $\kappa$-Minkowski spacetime is associated to a curved momentum space. The metric for 
 this curved momentum space is like to Eq.~(\ref{KPionMetric}). However, we can find many different metrics by using other methods such as embedding in 5-dimensional metric preserving groups, Riemannian hyperbolic momentum space, and two-time hyperbolic space \cite{Lizzi:2020tci}. 

As seen the constructions of the curved momentum space are not unique, several inequivalent momentum space geometries can be introduced. The curvature that we are discussing here has originated from the non-linearity in the combinations of the momentums as is given in relative locality theory.

\bibliography{main}

\end{document}